# Tracing scientist's research trends realtimely

Xianwen Wang*[1,2], Zhi Wang[†,1,2], Shenmeng Xu[†,1,2]

*1. WISE Lab, Faculty of Humanities and Social Sciences, Dalian University of Technology, Dalian 116085, China*

*2. School of Public Administration and Law, Dalian University of Technology, Dalian 116085, China*

\* Corresponding author. *Email address*: xianwenwang@dlut.edu.cn

† These authors contributed equally as second authors.

## Abstract

In this research, we propose a method to trace scientists' research trends realtimely. By monitoring the downloads of scientific articles in the journal of *Scientometrics* for 744 hours, namely one month, we investigate the download statistics. Then we aggregate the keywords in these downloaded research papers, and analyze the trends of article downloading and keyword downloading. Furthermore, taking both the download of keywords and articles into consideration, we design a method to detect the emerging research trends. We find that in scientometrics field, social media, new indices to quantify scientific productivity (g-index), webometrics, semantic, text mining, open access are emerging fields that scientometrics researchers are focusing on.

## Keywords

*Research trend; Altmetrics; Springer; Realtime; Scientometrics; Download*

## Introduction

Tracing research trends is one of the subjects which are of particular interest to scientists, because it helps them to grasp the realtime development and future direction of science and technology.

As the scientific community grows, academic publications are also increasing explosively, reaching an unprecedented number and involving more academic sectors and disciplines. Preferentially reading articles from specific journals can no longer satisfy the need of scientists to follow up the latest research trends. As a result, scholars today are increasingly interested in methods that can help them find hot topics in their specific scientific fields. Good filters for quality, importance, and relevance are necessary in the advance-phase preparation in academic researches (Neylon and Wu 2009), instead of the highly subjective selections before.

As the first step in the advance-phase preparation, reviewing literatures requires searching and downloading first. A series of research done by Kurtz et al. show that the way researchers access and read their technical literature has gone through a revolutionary change. "Whereas fifteen years ago nearly all use was mediated by a paper copy, today nearly all use is mediated by an electronic copy" (Kurtz and Bollen 2010). Accordingly, scientists read extensive literature when doing research, and the articles they

http://dx.doi.org/10.1007/s11192-012-0884-5

read are obtained by downloading from various science indexes and database. Articles being downloaded can reflect the research focus concerned by many scientists, because scientists download articles that they are interested in. The necessity of downloading makes it full-scale to study the research trends by investigating the downloads.

In addition, since there is a definite relationship between an article and its authors, it is viable to know about the leading-edge research by paying close attention to the leading scientists in that field. This evaluation can be achieved by measuring and analyzing the downloads of scientific papers. Meanwhile, scientists are also concerned about their own academy impact and whether their work is drawing colleagues' attention. So studying about the downloads helps them to identify themselves.

Previous studies have proposed two ways to analyze the research trends. The more direct but heavy and complicated way is to collect and read plenty of literatures, review them, and summarize the trends and directions for further research. Bibliometric methods, however, conduct statistical analysis of publication outputs of countries, research institutes, journals, and research fields (Cole 1989; Zitt & Bassecoulard 1994; Braun et al. 1995; Braun et al. 2000; Ding et al. 2001; Keiser & Utzinger 2005; Xie 2008), such as word frequency analysis, citation analysis, co-word analysis, etc. Reviewing related research about mining the hot topics and tracing scientists' research trends, various methods are being proposed on the basis of citations, number of publications, and other text-based data. Information such as source title, author keyword, keyword plus, and abstracts are also introduced in study of the research trend (Arrue & Lopez 1991; Qin 2000; Li et al. 2009).

Nevertheless, it is defective to evaluate the research trends just using traditional methods and just depending on information in formerly published scientific outputs.

Take citation analysis for example, there are several reasons. First of all, the publication of a scientific paper requires months to execute the review process, and as a result, significant publication delay will cause citation delay, and thus cause delay in the current research trend analysis. Second, as is known, there may be impact but certainly not citations. When an article provides scholars with inspirations and ideas that are not capable to directly support the research, it will not be cited, which doesn't mean it does not scholarly affect the author and the whole research trends. Sometimes, intentionally or not, even articles with strong and direct influence are not cited. These situations cannot be assessed. Thirdly, it is parochial to regard impact just as citations, since some influential theories, such as the Merton Miller theorem and Mendelian genetics, are widely accepted but seldom cited. A study examined articles in biogeography and found that only specific types of the influence is cited, and work that is "uncited" and "seldom cited" is used extensively. This study show that biogeographical scientists rely heavily on extremely large databases compiled by thousands of individuals over centuries in their research. However, there is "a generally accepted protocol by which authors provide substantial information about the databases they use, but they do not cite them" (MacRoberts and MacRoberts 2010). Moreover, Shuai et al. (2012) suggested that it is not always true that citation data represent an explicit, objective expression of impact by scientists.

In additon, an inevitable limitation maybe that valid academic writing is not only constituted with academic articles formally published in traditional journals. Many articles published in social media may have scientific influence or potential scientific influence, which cannot be easily evaluated. However, it is difficult to judge whether an article in a blog or a tweet is mature enough to be regarded as a scientific one. According to traditional forms of scholarly production, articles or other publications posted on web-based social media are not recognized as academic products (Lovink 2008; Borgman 2007; Kirkup 2010). Kirkup (2010) also suggested that these articles might be less problematic for students than traditional scientific papers, but "has been less enthusiastically embraced as offering alternatives for scholars and researchers".

Recently, realizing that increasing scholarly use of Web 2.0 tools presents an opportunity to create new filters, research into "altmerics" is receiving more and more attention (Priem et al. 2010). "Altmetrics is the creation and study of new metrics based on the Social Web for analyzing and informing scholarship." A diverse set of web-based social media like CiteULike, Mendeley, Twitter, and blogs now can be analyzed to inform real-time article recommendation and research trends. These metrics under the banner of "altmetrics" are based on social sources, and could yield broader, richer, and timelier assessments of current and potential scholarly impact (Koblenz 2011).

By now, many publishing groups offer evaluated tools for altmetrics. *Realtime tool* in *Springer*, *Altmetric APP* and *Mostdownloaded APP* in *Elsevier* are good examples. In addition, some journals and organizations provide instant analysis results of altmetrics, such as *Article-Level Metrics* (http://www.jmir.org/stats/overview) in *Journal of Medical Internet Research*, *Top Downloaded Articles* (http://www.stemcells.com/view/0/topdownloaded.html) in *Stem Cells*, *Download statistics* (http://discovery.ucl.ac.uk/past-statistics.html) in *UCL Discovery*, and *PLoS Impact Explorer* in *PloS* (http://altmetric.com/demos/plos.html), etc.

For example, *Springer* provides a function namely *Most Downloaded Articles* for every journal, which displays top five most downloaded articles from the journal during the past 7/30/90 days. Here we capture the *Most Downloaded Articles* from the website of *Scientometrics* journal at 8:20 on March 29, 2012 (Greenwich Mean Time). As Fig. 1 shows.

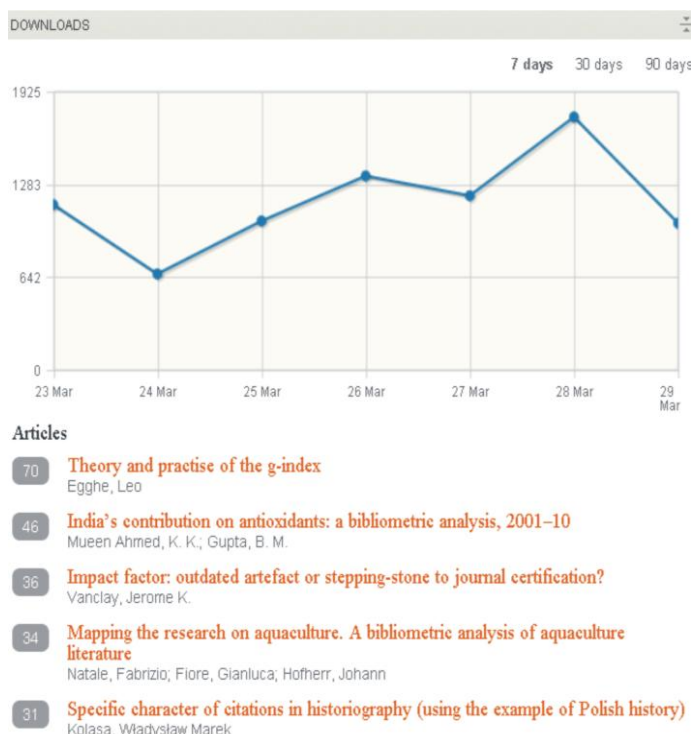

Fig. 1   Download statistics from *Scientometrics*

*Realtime tool* in *Springer* also provides keywords download statistics, as is shown in Fig. 2. Nevertheless, the tag cloud of the keywords has some drawbacks. First of all, the statistics cover all the papers and keywords in Springer, not by fields. However, most scientists are more interested in their own research areas. They rarely pay attention to and hardly understand the keywords in other areas. Secondly, the tag cloud includes only papers with keywords statistics, but many papers published in the 20$^{th}$ century do not have keywords, which means the keywords statistics of the tag cloud are incomplete.

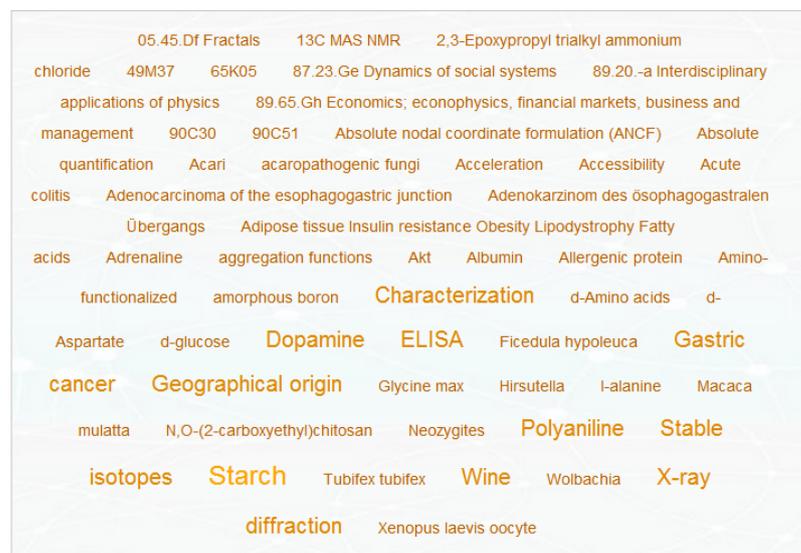

Fig. 2 Keywords download statistics from Springer (http://realtime.springer.com/keywords)

Recent efforts have explored the use of social networking on scholarly practice (Greenhow 2009; Veletsianos and Kimmons 2012). Kirkup (2010) investigated the function of blogging in academic practice and its contribution to academic identity and argued that academic blogging "offers the potential of a new genre of accessible academic production". Groth and Gurney (2010) analyzed the bibliometric properties of academic chemistry blogs and show the practical potential of this approach. Kjellberg (2011) described interviews with 12 researchers on their use and authoring of blogs.

As a microblogging platform, Twitter could offer faster, broader, and more nuanced metrics of scholarly communication to supplement traditional citation analysis (Priem and Costello 2010). Priem and Hemminger (2010) call for investigation into Twitter citations as part of a "scientometrics 2.0" that mines social media for new signals of scholarly impact. Weller and Puschmann (2011) explored the ways in which scholars use Twitter and related platforms to cite scientific articles. Other research examined how scientists use Twitter during conferences by analyzing tweets containing conference hashtags (Ebner and Reinhardt 2009; Letierce et al. 2010; Well et al. 2011).

Nevertheless, despite the growing speculation and early exploratory investigation into altmetrics, they mainly focus on the measurement of scientists' personal influence. In this study, however, we find scientists' hot topics and trace the research trends through altmetrics. Moreover, different from the previous studies, we pay attention to the downloads, because the articles which attracts scientists' attention will surely be downloaded to read but not necessarily be shared in Mendeley or discussed in Twitter.

We measure the research trends in scientometrics by analyzing the articles downloaded daily, weekly and monthly in the journal *Scientometrics*. We aggregate the keywords to go deep into the result. In fact, metrics are interlinked In general. Studies have shown that downloads statistics are in correlation with citation statistics and thus can predict future citation impact (Moed 2005; Brody et al. 2006; Jahandideh & Abdolmaleki 2007; O'Leary 2008), which is in line with our study.

## Data and methods

As is mentioned above, the necessity of downloading makes it full-scale to study the research trends by investigating the downloads.

Since December 2010, in order to "provide the scientific community with valuable information about how the literature is being used *right now*" (http://realtime.springer.com/about), Springer has launched a new free analytics tool, namely *realtime.springer.com*. It aggregates downloads of Springer journal articles and book chapters in real time from all over the world and displays the downloads in four visualization ways. The *map* shows which city the downloads are coming from, and the *Realtime Feed* displays constantly updating latest downloaded items, including the title, the source publication, authors, etc.

We conducted a series of studies using this tool, including the study on scientists' working timetable according the downloads map (Wang et al. 2012). In this study, we try to summarize the hot topics and research trends of the scientometrics field according to the downloaded articles. Here the journal *Scientometrics* is selected to be our research object. Three kinds of data need to be collected, namely the realtime downloading data, WoS data and Online First data.

### Realtime Downloading Data

We have been monitoring the realtime download statistics from the website of *realtime.springer.com* for a whole month. As Fig. 3 shows. From March 1 to March 31 2012, we record the time (Greenwich time), title, authors, Digital Object Identifier (DOI) of every item downloaded from *Scientometrics* round the clock.

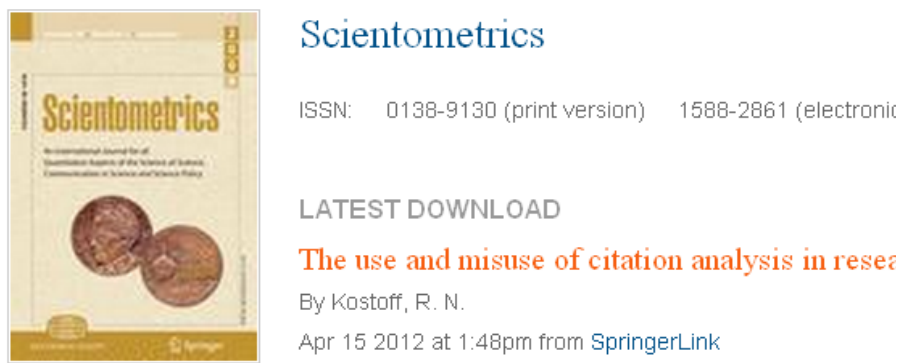

Fig. 3  Latest download of *Scientometrics* articles

### WoS Data

The WoS data is harvested from *webofknowledge.com*, on which the keywords information is provided. In total, 3172 records indexed in Web of Science from 1978 (Volume 1, Issue 1) to March 2012 (Volume 90, Issue 3) are collected. The majority of the data are labeled with DOI (Digital Object Identifier). For the 211 items without DOI, we check the original papers to complete this field.

Among the 3172 records, 503 items have DE field (descriptors, keywords given by authors), and 1780 records have ID field (Identifiers, added in Web of Science). Some items have both the DE field and ID field, and 1342 records have neither of them. For these 1342 items, we make word segmentation according to the titles. Other processes have also been conducted, such as plurality unifying, synonyms merging, etc.

### Online First Data

Since the new accepted articles before print publication have not been indexed in Web of Science, they need to be collected from the website of the journal, http://www.springerlink.com/content/101080.

### Methods

After the data processing, data are imported into the designed SQL Server database, as Fig. 4 shows. Three kinds of data are connected by the DOI as the primary key in the database.

From the realtimely downloaded data, we make statistical analysis for most downloaded articles. Linking with WoS data through DOI, we get most downloaded WoS papers. Nevertheless, for those Online First data, because they are just freshly published online, the downloading cannot be attributed to the intentional searching by scientists. Scientists who browse the website of Scientometrics regularly or are linked with RSS feeds are more likely to download online first articles which are not necessarily related to their current research and interests. Therefore, these downloads cannot fairly reveal the real research trends. In other words, these data would cause bias in our study, so a relatively low weight should be set on this portion of data to eliminate the bias. As a result, to simplify the research, we set the weight of Online First data as 0.

According to the keywords information from WoS data, we aggregate the most downloaded articles to most downloaded keywords. And then, we analyze the data at 3 levels, which are daily level, weekly level and month level analysis.

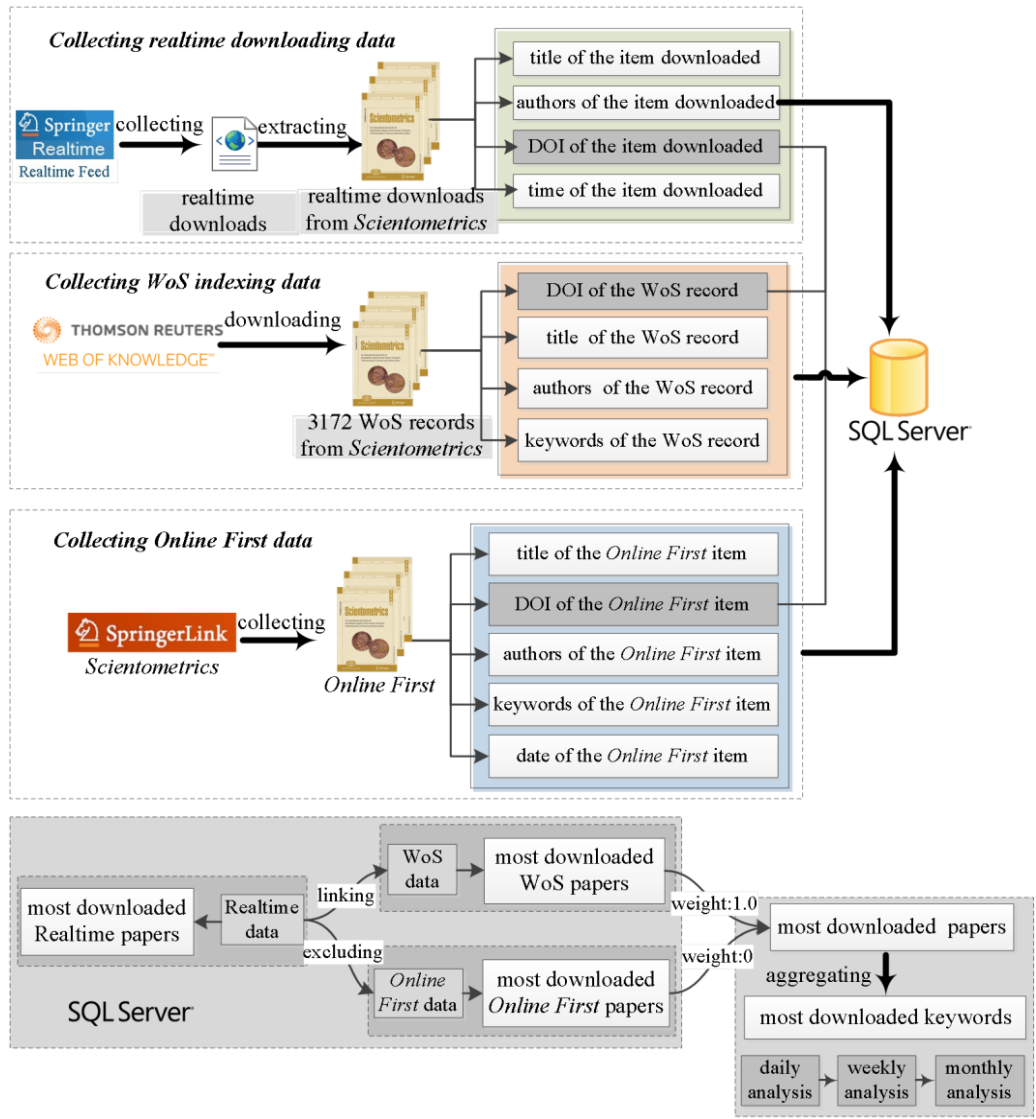

Fig. 4   Research framework

## Results

### Daily Downloads

Fig. 5 describes the number of downloads among the 31 days of this March. We can see that downloads in most of the weekdays are around 1000, while in the weekends, they significantly decrease, varying from 400 to 800. The red square dots denote the article downloads on weekends.

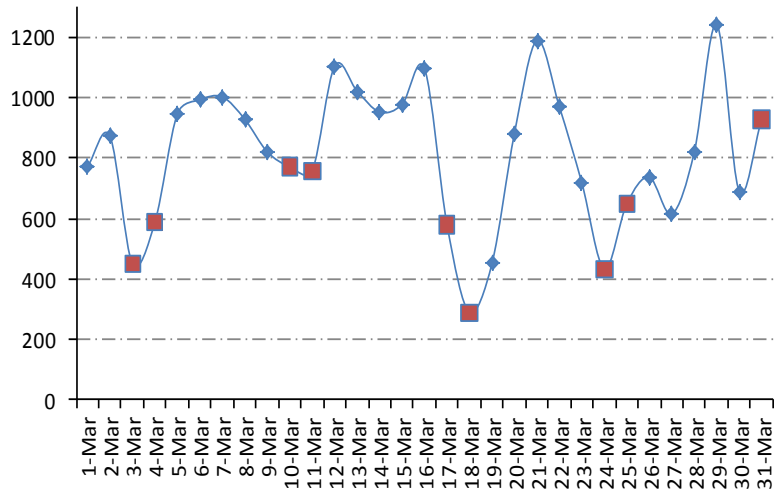

Fig. 5    Daily downloads of articles

**Most Downloaded Articles**

In Table 1, the top downloaded articles in the whole month of March are listed. These 21 articles are all downloaded more than 40 times, among which the top one is *Explicitly searching for useful inventions: dynamic relatedness and the costs of connecting versus synthesizing*, which was downloaded for 120 times. Moreover, *Theory and practise of the g-index* was downloaded 83 times and *Specific character of citations in historiography* 75 times.

Table 1    Most downloaded articles in March 2012

| title | downloads |
| --- | --- |
| Explicitly searching for useful inventions: dynamic relatedness and the costs of connecting versus synthesizing | 120 |
| Theory and practise of the g-index | 83 |
| Specific character of citations in historiography (using the example of Polish history) | 75 |
| Mapping the research on aquaculture. A bibliometric analysis of aquaculture literature | 74 |
| Weighted indices for evaluating the quality of research with multiple authorship | 72 |
| Software survey: VOSviewer, a computer program for bibliometric mapping | 62 |
| Funding acknowledgement analysis: an enhanced tool to investigate research sponsorship impacts: the case of Nanotechnology | 59 |
| Mapping the (in)visible college(s) in the field of entrepreneurship | 57 |
| Negative results are disappearing from most disciplines and countries | 55 |
| Network model of knowledge diffusion | 54 |
| Research on the semantic-based co-word analysis | 51 |
| Using author co-citation analysis to examine the intellectual structure of e-learning: A MIS perspective | 48 |
| Scientific collaboration in Library and Information Science viewed through the Web of Knowledge: the Spanish case | 48 |
| The organization of scientific knowledge: the structural characteristics of keyword networks | 46 |
| Bibliometric trend analysis on global graphene research | 45 |

http://dx.doi.org/10.1007/s11192-012-0884-5

| | |
|---|---|
| Using social media data to explore communication processes within South Korean online innovation communities | 44 |
| Agent-based computing from multi-agent systems to agent-based models: a visual survey | 43 |
| The Triple Helix of university-industry-government relations | 43 |
| Co-citation analysis and the search for invisible colleges: A methodological evaluation | 41 |
| The blockbuster hypothesis: influencing the boundaries of knowledge | 41 |
| Sources of Google Scholar citations outside the Science Citation Index: A comparison between four science disciplines | 41 |

**Most Downloaded Keywords**

We analyze the top articles in every week, and aggregate them to keywords statistics. As is shown in Table 2, for the four one-week periods, the top 5 most downloaded keywords are mostly similar, including *science, citation, indicator, bibliometrics, citation analysis*. These stable words are among the most frequently used words in the field of scientometrics. Besides, words like science and indicator, whose characteristics are relatively week, are commonly used in scientific papers in other research fields.

Nevertheless, significant features are shown in these downloaded keywords, because some of them are of great volatility. Take *patent* for example. During week 1 (from March 1 to March 7), it was downloaded 202 times, ranking 10th; during week 2 (from March 8 to March 14), it was downloaded only 110 times, ranking 24th; during week 3 (from March 15 to March 21), the downloaded times furthered down to only 89 times; and during week 4 (from March 22 to March 28), the curve rise again to 109. For another keyword *impact factor*, the downloaded times and ranks during the four weeks are 146 (17), 185 (13), 185 (11) and 151 (14).

Table 2   Most downloaded keywords in March 2012

| week1 | | week2 | | week3 | | week4 | |
|---|---|---|---|---|---|---|---|
| keywords | times | keywords | times | keywords | times | keywords | times |
| science | 694 | science | 837 | science | 693 | science | 682 |
| citation | 397 | Indicator | 520 | Citation | 393 | indicator | 408 |
| indicator | 357 | citation | 452 | indicator | 375 | citation | 378 |
| bibliometrics | 330 | bibliometrics | 370 | bibliometrics | 367 | citation analysis | 296 |
| citation analysis | 280 | Journal | 325 | journal | 302 | bibliometrics | 265 |
| journal | 251 | citation analysis | 324 | citation analysis | 265 | journal | 256 |
| h-index | 217 | Impact | 310 | h-index | 252 | impact | 221 |
| publication | 207 | h-index | 266 | impact | 231 | h-index | 217 |
| impact | 202 | university | 239 | collaboration | 219 | collaboration | 193 |
| ***patent*** | ***202*** | publication | 238 | publication | 202 | innovation | 189 |
| innovation | 181 | collaboration | 238 | **impact factor** | **185** | technology | 175 |
| university | 170 | scientometrics | 213 | university | 165 | pattern | 164 |
| co-authorship | 168 | **impact factor** | **185** | scientometrics | 156 | publication | 163 |
| collaboration | 167 | ranking | 178 | innovation | 154 | **impact factor** | **151** |
| scientometrics | 160 | technology | 178 | ranking | 148 | scientometrics | 150 |
| technology | 157 | innovation | 158 | research performance | 137 | ranking | 146 |
| **impact factor** | **146** | pattern | 150 | co-authorship | 135 | research performance | 145 |
| bibliometrics analysis | 144 | country | 147 | technology | 123 | university | 141 |

http://dx.doi.org/10.1007/s11192-012-0884-5

| | | | | | | | |
|---|---|---|---|---|---|---|---|
| research performance | 140 | co-authorship | 145 | pattern | 116 | nanotechnology | 130 |
| nanotechnology | 140 | research performance | 138 | bibliometrics analysis | 115 | bibliometrics indicator | 116 |
| ranking | 130 | network | 136 | productivity | 115 | triple helix | 112 |
| linkage | 129 | bibliometrics indicator | 128 | model | 110 | co-authorship | 110 |
| pattern | 126 | bibliometrics analysis | 121 | network | 109 | ***patent*** | ***109*** |
| search | 106 | ***patent*** | ***110*** | nanotechnology | 109 | productivity | 99 |
| network | 105 | china | 108 | bibliometrics indicator | 107 | scientific collaboration | 97 |
| triple helix | 105 | scientific collaboration | 105 | quality | 102 | network | 94 |
| research collaboration | 101 | quality | 101 | triple helix | 99 | co-citation | 93 |
| bibliometrics indicator | 100 | model | 101 | country | 97 | quality | 88 |
| performance | 98 | nanotechnology | 100 | scientific collaboration | 96 | knowledge | 83 |
| china | 97 | performance | 95 | ***patent*** | ***89*** | scientific literature | 83 |

Accordingly, we calculate the keywords download ratio, which can be expressed by the weekly downloads divided by the total number of downloads.

$$Ratio1 = \frac{downloads\ of\ the\ keyword}{total\ downloads}$$

Fig. 6 reveals the variation of six keywords. On one hand, during week 1, the ratio of downloads of *patent* is about 8.1%. It slipped to 5.9% and furthered down to 5.7% in week 2 and week 3 correspondingly. During week 4, however, the ratio rose to 6.5% again. For the keyword *h-index*, the download ratio increased slightly from 4.5% in week 1 to 5.1% in week 3, and dropped to 4.7% in week 4. The keyword *impact factor* changes consistently with *patent*. On the other hand, for the other three keywords, which are *mapping, peer review,* and *co-word analysis*, their download ratios are stable in these 4 weeks.

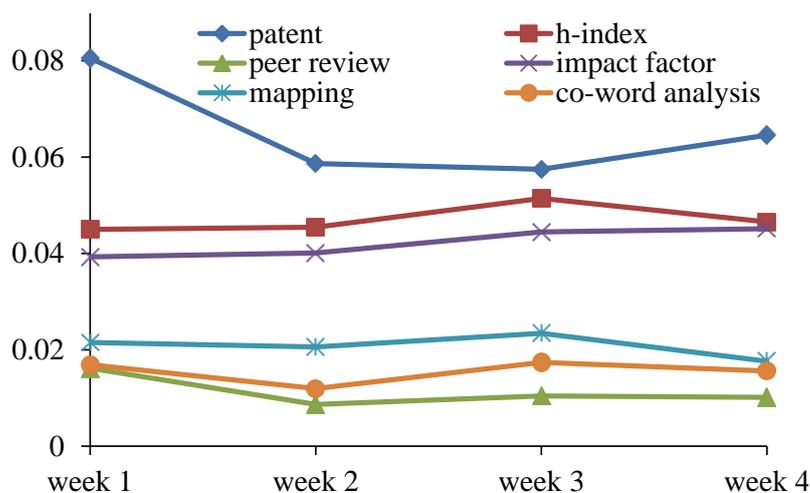

Fig. 6   Weekly fluctuation of the ratio of keywords downloads

**Emerging Research Trends Analysis**

In the relatively mature scientific fields, due to the long history of the research area and the great quantity of scientific articles, the downloads and download ratios of keywords would be relatively high. Examples are the keywords *citation, bibliometrics, co-authorship*, etc.

We calculate the ratio of keywords downloads to published articles as follow.

$$Ratio2 = \frac{downloads\ of\ keyword}{number\ of\ papers\ have\ the\ keyword}$$

For example, the downloads of keyword *citation* is 4214, and the number of published articles in *Scientometrics* which have *citation* as keyword is 433, then the calculated result of this ratio is about 9.73.

In those emerging research fields, due to the relatively short history, there is not much published articles. As a result, keywords in these articles are seldom downloaded. However, if we divide the keywords downloads by the number of articles that has it as a keyword, it would be interesting. For example, there are only 3 articles published in *Scientometrics* which have the keyword *twitter*, but the downloads of keyword *twitter* reaches 123 in March 2012. Therefore, the ratio for twitter to articles is as high as 41.

Consequently, we design a method to trace the emerging research trends.

(1)    The keyword is new in recent years or in specific scientific journal/ field.

(2)    The keyword downloads is relatively high. Here we set the criterion as 50.

(3)    The ratio of keyword downloads to published articles is greater than 20.

50 most downloaded keywords are selected for our analysis. We calculated the ratio, and the results are displayed in Fig. 7. In this scatter plot, each dot stands for a keyword. The horizontal axis is the number of published articles which have the keyword, while the vertical axis is the ratio of keyword downloads to published articles. Dots located at the upper left corner of the scatter plot have the ratio greater than 20. As is seen from the figure, some research trends can be revealed. *Twitter* reflects the rapid development of altmetrics based on social media networks. *G-index*, which was proposed by Leo Egghe in 2006, are also attracting scientometrics scientists' interests. *Vosviewer* is a new visualization software developed by CWTS Leiden University in 2009, which has received much attention since its release. Other keywords, including *webometrics, latent semantic, open access, etc.,* all reveal recent research trends in scientometrics.

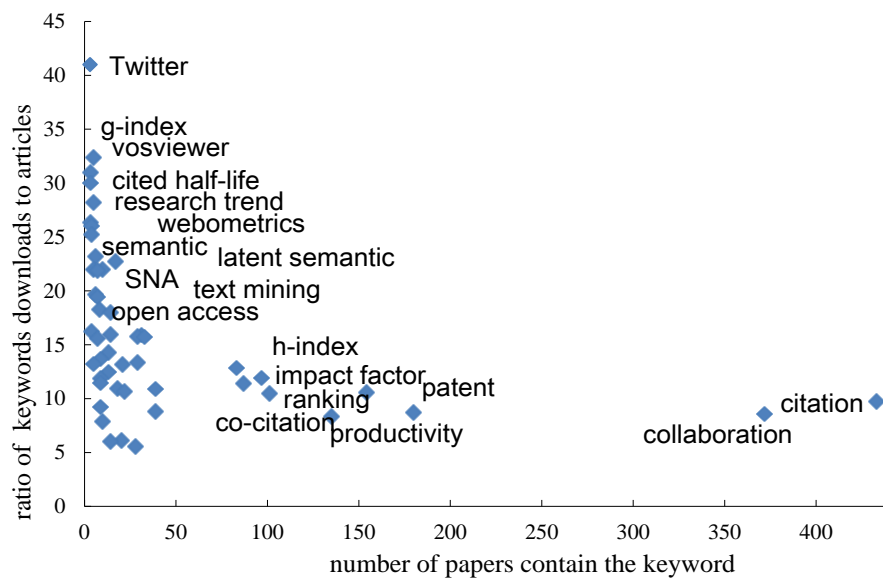

Fig. 7   Ratio of keywords downloads to published articles

## Conclusions and discussion

In this research, we propose a method to trace scientists' research trends realtimely. We monitor the downloads of scientific articles in *Scientometrics* for one whole month, and dig deep into the download statistics. By building a large database and aggregating the keywords in these articles, the trends of article downloading and keyword downloading are revealed, which can finely indicate the research trends because when scientists read literatures, they choose articles that they are interested in, and the articles are necessarily obtained by downloading from science indexes and databases.

Furthermore, meaningful indicators are designed to detect the emerging research trends. Taking both the downloads and publications of articles into consideration, we design a method to track the changes and to identify the newer and "hotter" research focus. We find that in Scientometrics field, social media, new indices to quantify scientific productivity (g-index), webometrics, semantic, text mining, and open access are emerging areas that information scientists are focusing on. These topics will be leading research trends in the near future.

Since a very small minority of papers may be downloaded involuntarily or for other irrelevant reasons, the arbitrary and randomness of downloading cannot be completely excluded. This figure is difficult to retrieve and measure, but in consideration of the low probability, we don't take it into account in this paper.

To find the relation between downloads and citations requires observation over a long period. In this article, we only analyze the data in one month, however, since March 1st 2012, we have been keeping recording the downloading data 24/7. After a longer period of monitoring and recording, using more realtime data, we will go deeper into this analysis in the future.

Wang, X. W., Wang, Z., Xu, S. M. (2012). Tracing scientist's research trends realtimely [J] *Scientometrics*, DOI: 10.1007/s11192-012-0884-5.

## Acknowledgments

The research is supported by the project of "Social Science Foundation of China"(Grant No. 10CZX011), the project of "Specialized Research Fund for the Doctoral Program of Higher Education of China" (Grant No. 2009041110001), as well as the project of "Fundamental Research Funds for the Central Universities" (Grant No. DUT12RW309).

http://dx.doi.org/10.1007/s11192-012-0884-5